\documentclass[12pt]{article}
\usepackage[margin=1in]{geometry}
\usepackage{amsmath}

\begin{document}

\newcommand{\z}[1]{\zeta({\bf k}_{#1})}
\newcommand{\zg}[1]{\zeta_1({\bf k}_{#1})}
\newcommand{\bk}{{\bf k}}

\thispagestyle{empty}
\begin{flushright}
\parbox[t]{1.8in}{USTC-ICTS-10-03}
\end{flushright}

\vspace*{0.2in}

\begin{center}
{\large \bf Quadra-Spectrum and Quint-Spectrum from \\[.1in] Inflation  and Curvaton Models}

\vspace*{0.45in} {Chunshan Lin $^{a,b,}$\footnote{lics@mail.ustc.edu.cn} and Yi Wang $^{b,}$\footnote{wangyi@hep.physics.mcgill.ca}}
\\[.15in]
{\em $^a$ Interdisciplinary Center of Theoretical Studies\\
University of Science and Technology of China\\
Hefei, Anhui 230026, China\\[.1in]
$^b$ Physics Department, McGill University,\\ Montreal, H3A2T8, Canada
 }
\end{center}
\vspace{.2in}

\begin{center}
{\bf Abstract}
\end{center}
\noindent

We calculate the quadra-spectrum and quint-spectrum, corresponding to five and six point correlation functions of the curvature perturbation. For single field inflation with standard kinetic term, the quadra-spectrum and quint-spectrum are small, which are suppressed by slow roll parameters. The calculation can be generalized to multiple fields. When there is no entropy perturbation, the quadra-spectrum and quint-spectrum are suppressed as well. With the presence of entropy perturbation, the quadra-spectrum and quint-spectrum can get boosted. We illustrate this boost in the multi-brid inflation model. For the curvaton scenario, the quadra-spectrum and quint-spectrum are also large in the small $r$ limit. We also calculate representative terms of quadra-spectrum and quint-spectrum for inflation with generalized kinetic terms, and estimate their order of magnitude for quasi-single field inflation.

\vfill
\newpage
\setcounter{page}{1}

\section{Introduction}

The non-Gaussian features in the cosmic microwave background radiation (CMB) and large scale structure (LSS) have become one of the central issues in modern cosmology. Theoretically, a large primordial non-Gaussianity indicates new physics beyond single field slow roll inflation. Moreover, one can extract a number of observables from the measurement of non-Gaussianity, for example, the size and shape for bispectrum and trispectrum respectively. These observables can help us greatly in distinguishing between different inflation models.

The three point correlation function, or the so called bispectrum has been relatively well investigated in the literature. Single field slow roll inflation has very small bispectrum \cite{Maldacena:2002vr}. Large bispectrum can be obtained from non-standard kinetic terms \cite{Chen:2006nt}, non-local inflation \cite{Barnaby:2007yb}, features in the Lagrangian \cite{Chen:2006xjb}, curvatons \cite{Lyth:2002my}, old curvatons \cite{Gong:2009dh}, multiple inflatons \cite{Vernizzi:2006ve}, quasi-single field inflation \cite{Chen:2009we}, multi-brid inflation \cite{Naruko:2008sq}, bifurcating trajcetores \cite{Li:2009sp}, thermal components \cite{Chen:2007gd} and so on. Experimentally, the WMAP7 bound \cite{Komatsu:2010fb} for the bispectral estimator $f_{NL}$ in local, equilateral and orthogonal shapes are given as (68\% CL)
\begin{equation}
  f_{NL}^{\rm local}=32\pm 21~, \quad f_{NL}^{\rm equil}=26\pm 140~,
  \quad f_{NL}^{\rm orthog}=-202\pm 104~.
\end{equation}
from which one finds that a Gaussian model is still consistent with data, however the central values of all the three shapes are far from zero. In the near future, the Planck mission is able to reduce the uncertainty to $\Delta f_{NL}^{\rm local}\simeq 5$. The error bars for bispectra of other shapes are also expected to be greatly reduced.

The four point correlation function, or the so called trispectrum, is also calculated recently in a number of models. Large trispectrum can show up in inflation with single field (for complete result, see \cite{Chen:2009bc}; for earlier work, see \cite{Huang:2006eha,Arroja:2008ga}) and multiple field \cite{Gao:2009at} inflation with generalized kinetic terms, curvaton scenario \cite{Sasaki:2006kq}, multi-brid inflation \cite{Huang:2009xa}, and so on. A (95\% CL) bound on trispectrum from the WMAP5 data is given as \cite{Smidt:2010sv}
\begin{equation}
  -3.80\times 10^{6}<g_{NL}<3.88\times 10^6~,\quad -3.2\times 10^{5}<\tau_{NL}<3.3\times 10^{5}.
\end{equation}
In the Planck mission, the uncertainty can be reduced to $\Delta \tau_{NL}\simeq 560$. The signal to noise ratio of $\tau_{NL}$ up to a given $l$ is shown to be \cite{Kogo:2006kh}
\begin{equation}\label{sn}
  (S/N)(<l)\sim 2.2\times 10^{-9} f_{NL}^2l^2~.
\end{equation}
Thus with a large $f_{NL}$, trispectrum is hopefully to be measured in Planck.

For higher point correlation functions than four point, the investigations are much less in the literature. Some results on the consistency relations \cite{Huang:2006eha, Li:2008gg}, and in $\delta N$ formalism \cite{Yokoyama:2008by} are available. However a detailed investigation is not yet present. In this paper, we shall calculate explicitly the five and six point correlation functions (the quadra-spectrum and the quint-spectrum) and extract the predictions in several models. In Section 2, we discuss the local shape of quadra-spectrum and quint-spectrum. We calculate the quadra-spectrum and quint-spectrum for single field and multiple field inflation models generically  in Section 3, for Multi-brid inflation in Section 4, and for the curvaton scenario in Section 5. We also calculate representative terms of equilateral quadra-spectrum and quint-spectrum in Section 6, using single field inflation with generalized kinetic terms. We conclude and estimate the order of magnitude for quadra-spectrum and quint-spectrum for quasi-single field inflation in Section 7.

\section{The local ansatz}

The most well studied non-Gaussian feature is the local shape non-Gaussianity. In the local ansatz, $\zeta(x)$ can be expanded locally at a spacetime point $x$ as
\begin{align}\label{zetalocal}
  \zeta(x)&=\zeta_1(x)+\frac{3}{5}f_{NL}\zeta_1^2(x)+\frac{9}{25}g_{NL}\zeta_1^3(x)+\frac{27}{125}h_{NL}\zeta_1^4(x)+\frac{81}{625}i_{NL}\zeta_1^5(x)+{\cal O}(\zeta_1^6)
  \nonumber\\
  &=\sum_{n=1}^{\infty}\frac{\zeta_n(x)}{n!}
  ~.
\end{align}
Thus $\zeta_i$'s can be written in terms of the non-Gaussian estimators as
\begin{gather}\label{relng}
  \zeta_2=\frac{6}{5}f_{NL}\zeta_1^2~, \qquad
  \zeta_3=\frac{54}{25}g_{NL}\zeta_1^3~,\qquad
  \zeta_4=\frac{648}{125}h_{NL}\zeta_1^4~,\qquad
  \zeta_5=\frac{1944}{125}i_{NL}\zeta_1^5~.
\end{gather}
The virtue of local non-Gaussianity is two folded: On the one hand, the local anzatz makes the analysis of non-Gaussian features simpler technically. On the other hand, in a number of known examples, non-Gaussianity is produced in super-Hubble scales. In this case, only the local shape non-Gaussianity is possible from causal considerations \footnote{By contract, non-Gaussianity may be produced in sub-Hubble scales, but get amplified in super-Hubble scales. In this case, the shape might not be local \cite{Li:2008jn}.}.

In the Fourier space, the local ansatz corresponds to
\begin{align}
  &\z{}=\zg{}+\frac{3}{5}f_{NL}\int\frac{d^3k_1}{(2\pi)^3}\zg{1}\zeta_1(\bk-\bk_1)
  +\frac{9}{25}g_{NL}\int\frac{d^3k_1}{(2\pi)^3}\int\frac{d^3k_2}{(2\pi)^3}
  \zg{1}\zg{2}\zeta_1(\bk-\bk_1-\bk_2)
  \nonumber\\
  &
  +\frac{27}{125}h_{NL}\int\frac{d^3k_1}{(2\pi)^3}\int\frac{d^3k_2}{(2\pi)^3}\int\frac{d^3k_3}{(2\pi)^3}\zg{1}\zg{2}\zg{3}\zeta_1(\bk-\bk_1-\bk_2-\bk_3)
  \nonumber\\
  &
  +\frac{81}{625}i_{NL}\int\frac{d^3k_1}{(2\pi)^3}\int\frac{d^3k_2}{(2\pi)^3}\int\frac{d^3k_3}{(2\pi)^3}\int\frac{d^3k_4}{(2\pi)^3}\zg{1}\zg{2}\zg{3}\zg{4}\zeta_1(\bk-\bk_1-\bk_2-\bk_3-\bk_4)
  \nonumber\\
  &
  +{\cal O}(\zeta_1^6)~.\label{fourier}
\end{align}

Following the standard definition, the shape functions of $\z{}$ up to four points are defined as
\begin{gather}
  \langle \z{1}\z{2} \rangle = (2\pi)^3 P(k_1)\delta^3(\sum_{n=1}^2 \bk_n)~, \nonumber\\
  \langle \z{1}\z{2}\z{3} \rangle = (2\pi)^3 B(\bk_1,\bk_2,\bk_3)\delta^3(\sum_{n=1}^3 \bk_n)~,\nonumber\\
  \langle \z{1}\z{2}\z{3}\z{4} \rangle = (2\pi)^3 T(\bk_1,\bk_2,\bk_3,\bk_4)\delta^3(\sum_{n=1}^4 \bk_n)~,\label{def234}
\end{gather}
where $B$ and $T$ denote the shape of non-Gaussianity, which are functions of the external momenta, as well as background quantities. Similarly, we define the shape functions for the five point correlation function and the six point correlation function as
\begin{gather}
  \label{def5}\langle \prod_{n=1}^5 \z{n} \rangle = (2\pi)^3 Q(\bk_1,\bk_2,\bk_3,\bk_4,\bk_5)\delta^3(\sum_{n=1}^5 \bk_n)~,\\
  \label{def6}\langle \prod_{n=1}^6 \z{n} \rangle = (2\pi)^3 U(\bk_1,\bk_2,\bk_3,\bk_4,\bk_5,\bk_6)\delta^3(\sum_{n=1}^6 \bk_n)~.
\end{gather}
In the local ansatz, the momenta dependence in the shape functions can be calculated explicitly, without knowing the detail mechanism how the correlation is established. Inserting Eq. \eqref{fourier} into Eqs. \eqref{def234}, \eqref{def5} and \eqref{def6}, we have
\begin{gather}\label{shapeB}
  B(\bk_1,\bk_2,\bk_3)=\frac{6}{5}f_{NL}
  \left[P(k_1)P(k_2)+2{~\rm perm.}\right]~.
  \\
  T(\bk_1,\bk_2,\bk_3,\bk_4)=
  \frac{54}{25}g_{NL}\left[ P(k_1)P(k_2)P(k_3) +3 {~\rm perm.} \right] \nonumber\\
  +\tau_{NL} \left[ P(k_1)P(k_2)P(|\bk_1+\bk_3|)+11{~\rm perm.}\right]~. \label{shapeT}
\end{gather}

\begin{align}
   Q(\bk_1,\bk_2,& \bk_3,\bk_4,\bk_5) =
  \frac{648}{125}h_{NL} \left[ P(k_1)P(k_2)P(k_3)P(k_4)+4{~\rm perm.} \right] \nonumber\\
  &+\theta_{NL} \left[ P(k_1)P(k_2)P(k_3)P(|\bk_1+\bk_4|)+59{~\rm perm.}\right] \nonumber\\
  &+\phi_{NL} \left[ P(k_1)P(k_2)P(|\bk_1+\bk_3|)P(|\bk_2+\bk_4|) +59{~\rm perm.} \right] ~.\label{shapeQ}
\end{align}
\begin{align}\label{shapeU}
 U(\bk_1,\bk_2,& \bk_3,\bk_4,\bk_5,\bk_6)=
  \frac{1944}{125}i_{NL}\left[ P(k_1)P(k_2)P(k_3)P(k_4)P(k_5)+5{~\rm perm.} \right] \\
  &+\alpha_{NL}\left[ P(k_1)P(k_2)P(k_3)P(k_4)P(|\bk_1+\bk_5|) +119{~\rm perm.} \right] \nonumber\\
  &+\beta_{NL} \left[P(k_1)P(k_2)P(k_3)P(k_4)P(|\bk_1+\bk_2+\bk_5|) + 89{~\rm perm.} \right]\nonumber\\
  &+\gamma_{NL}\left[ P(k_1)P(k_2)P(k_3)P(|\bk_1+\bk_4|)P(|\bk_2+\bk_5|)+359 {~\rm perm.}\right]\nonumber\\
  &+\delta_{NL}\left[ P(k_1)P(k_2)P(k_3)P(|\bk_1+\bk_4|)P(|\bk_1+\bk_4+\bk_5|)+359 {~\rm perm.}\right]\nonumber\\
  &+\lambda_{NL} \left[ P(k_1)P(k_2)P(|\bk_1+\bk_3|)P(|\bk_2+\bk_4|)P(|\bk_1+\bk_3+\bk_5|)+ 359{~\rm perm.} \right]\nonumber
  ~,
\end{align}
where
\begin{gather}
  \tau_{NL}=\frac{36}{25}f_{NL}^2~,
  \nonumber\\
  \theta_{NL}=\frac{324}{125}f_{NL}g_{NL}~,  \quad \phi_{NL}=\frac{216}{125}f_{NL}^3~, \nonumber\\
  \alpha_{NL}=\frac{3888}{625}f_{NL}h_{NL}~, \quad \beta_{NL}=\frac{2916}{625}g_{NL}^2~, \nonumber\\
  \gamma_{NL}=\delta_{NL}=\frac{1944}{625}f_{NL}^2g_{NL}~, \quad \lambda_{NL}=\frac{1296}{625}f_{NL}^4~. \label{localid}
\end{gather}
In the above calculation, we are considering only connected parts of the correlation functions. The disconnected parts are automatically dropped. The function $Q$, as well as some results in terms of the $\delta N$ formalism, are also obtained in \cite{Yokoyama:2008by}.

One should also note that Eq. \eqref{localid} only holds when the local expansion \eqref{zetalocal} is valid. In some generalized cases (for example, multi-field inflation and multi-brid inflation, which we shall show below), Eq. \eqref{zetalocal} no longer holds. Instead, the nonlinearity is expressed as a product of different fields at the same spatial point. In this case, the shape functions \eqref{shapeB}, \eqref{shapeT}, \eqref{shapeQ} and \eqref{shapeU} are still the same. However, the coefficients $\tau_{NL}$, $\theta_{NL}$, $\phi_{NL}$, $\alpha_{NL}$, $\beta_{NL}$, $\gamma_{NL}$, $\delta_{NL}$ and $\lambda_{NL}$ should be calculated independently. Eq. \eqref{localid} can not be used in this case.

\section{Single field and multi-field inflation}

The $\delta N$ formalism \cite{deltaN} is a powerful tool to calculate cosmic perturbations. Especially, it is very convenient to calculate local non-Gaussianity using the $\delta N$ formalism \cite{Lyth:2005fi}. The curvature perturbation is related with the perturbation of the inflaton field as
\begin{equation}
  \zeta(x)=N'\delta\phi_*+\frac{N''}{2}\delta\phi_*^2+\frac{N'''}{6}\delta\phi_*^3+\frac{N''''}{24}\delta\phi_*^4+\frac{N'''''}{120}\delta\phi_*^5+{\cal O}(\delta\phi_*^6)~,
\end{equation}
where $\delta\phi_*$ is the Hubble crossing value of the inflaton fluctuation in the spatial flat gauge, and prime is derivative with respect to $\phi_*$. In the current section, we assume that nonlinearity is mainly developed on super-Hubble scales, which is verified in bispectrum and trispectrum. This assumption corresponds to assuming
\begin{equation}
  \delta\phi_*=\delta_1\phi_{*}~,
\end{equation}
without non-Gaussian components. In this case, the non-Gaussian estimators can be directly identified with the derivatives of e-folding number as
\begin{gather}
  f_{NL}=\frac{5N''}{6N'^2}~,\quad g_{NL}=\frac{25N'''}{54N'^3}~,\quad h_{NL}=\frac{125N''''}{648N'^4}~,\quad i_{NL}=\frac{125N'''''}{1944N'^5}~.
\end{gather}
Other estimators are given in Eq. \eqref{localid}. One note from this result that as one goes to higher order, the results are suppressed by more powers of slow roll parameters. Here and afterwards, we only calculate the leading order non-Gaussianity, and neglected the contribution from loop diagrams, which are suppressed by at least of order $10^{-9}$, thus not likely to be observable.

Note that by saying the above result is suppressed by slow roll parameters, we assume the time derivatives of the slow roll parameters are also small. On the other hand, if the time derivatives are large, for example, in the case that the inflaton field has small oscillation during the slow roll, the quadra-spectrum and quint-spectrum can get amplified, by powers of the oscillation frequency divided by the Hubble parameter.

With the assumption that the sub-Hubble correlations can be neglected, the above calculation is also straightforward to generalize into multi-field case, with an additional assumption that isocurvature perturbation is absent (we shall go beyond this assumption in the next section). In multiple field inflation, $\delta N$ is expanded as
\begin{equation}\label{mdn}
  \zeta(x)=\sum_i N_i\delta\phi_{*i}+\frac{1}{2}\sum_{ij}N_{ij}\delta\phi_{*i}\delta\phi_{*j}+ \cdots~.
\end{equation}
By this expansion, we have assumed that the e-folding number difference depend only on the fields at horizon crossing. In other words, the uniform total energy density slice is the same as the uniform energy density slice for each component fields. This corresponds to the assumption of absence of entropy perturbation. Otherwise, one should express the fields on the decay hypersurface of the last field (or other hypersurface where isocurvature perturbation vanishes) as functions of the fields at the Hubble crossing time before doing the variation of $\phi_{*i}$. The estimators $f_{NL}$, $g_{NL}$, $h_{NL}$ and $i_{NL}$ can be written as
\begin{gather}
  f_{NL}=\frac{5\sum_{ij}N_{ij}N_iN_j}{6\left(\sum_i N_i^2\right)^2}~,\qquad g_{NL}=\frac{25\sum_{ijk}N_{ijk}N_iN_jN_k}{54\left(\sum_i N_i^2\right)^3}~,
  \nonumber\\
  h_{NL}=\frac{125\sum_{ijkl}N_{ijkl}N_{i}N_{j}N_{k}N_{l}}{648\left(\sum_{i}N_i^2\right)^4}~, \quad
  i_{NL}=\frac{125\sum_{ijklm}N_{ijklm}N_{i}N_{j}N_{k}N_{l}N_{m}}{1944\left( \sum_i N_i^2 \right)^5} ~.
\end{gather}
As Eq. \eqref{mdn} is different from Eq. \eqref{zetalocal}, the non-Gaussian estimators $\theta_{NL}, ~\phi_{NL}, ~\alpha_{NL}, ~\beta_{NL}, ~\gamma_{NL}$, $\delta_{NL}$ and $\lambda_{NL}$ are also no longer given by Eq. \eqref{localid}. Instead, after taking expectation values, one can find
\begin{gather}
  \tau_{NL}=\frac{\sum_{ijk}N_{ij}N_{ik}N_jN_k}{\left(\sum_i N_i^2\right)^3}~, \nonumber\\
  \theta_{NL}=\frac{\sum_{ijkl}N_{ijk}N_{il}N_jN_kN_l}{\left(\sum_i N_i^2\right)^4} ~,
  \quad
  \phi_{NL}=\frac{\sum_{ijkl}N_{ij}N_{ik}N_{jl}N_kN_l}{\left(\sum_i N_i^2\right)^4} ~,
  \nonumber\\
  \alpha_{NL}=\frac{\sum_{ijklm}N_{ijkl}N_{im}N_jN_kN_lN_m}{\left(\sum_i N_i^2\right)^5}~,
  \quad
  \beta_{NL}=\frac{\sum_{ijklm}N_{ijk}N_{ilm}N_jN_kN_lN_m}{\left(\sum_i N_i^2\right)^5} ~,
  \nonumber\\
  \gamma_{NL}=\frac{\sum_{ijklm}N_{ijk}N_{il}N_{jm}N_kN_lN_m}{\left(\sum_i N_i^2\right)^5}~,
  \quad
  \delta_{NL}=\frac{\sum_{ijklm}N_{ijk}N_{il}N_{lm}N_jN_kN_m}{\left(\sum_i N_i^2\right)^5} ~,
  \nonumber\\
  \lambda_{NL}=\frac{\sum_{ijklm}N_{ij}N_{kl}N_{im}N_{km}N_jN_l}{\left(\sum_i N_i^2\right)^5}~. \label{deltaNmpl}
\end{gather}
Note that in multi-field inflation, all the relevant fields are assumed to be light. Thus the resulting non-Gaussianity is typically suppressed by slow roll parameters. It is interesting to see whether loop corrections \cite{Cogollo:2008bi}, or a hierarchy of different slow roll parameters could amplify the non-Gaussianities. But these aspects are beyond the scope of the current work.

As noticed in \cite{Suyama:2007bg}, using Cauchy-Schwarz inequality
\begin{equation}
  \sum_{ij} (A_iA_i) (B_jB_j) \geq \sum_{ij} (A_i B_i) (A_j B_j)~,
\end{equation}
where $A_i$ and $B_j$ are real vectors, one can derive
\begin{equation}
  \tau_{NL}\geq \frac{36}{25}f_{NL}^2~.
\end{equation}
Similarly, using Cauchy-Schwarz inequality once, we have
\begin{equation}
  \beta_{NL}\geq \frac{2916}{625}g_{NL}^2~.
\end{equation}
Using Cauchy-Schwarz inequality twice, we have
\begin{equation}
  \lambda_{NL} \geq \frac{1296}{625}f_{NL}^4~.
\end{equation}
For other equations in Eqs.\eqref{localid}, one can not find similar inequalities correspondingly. Instead, we have checked that similar inequalities for $\theta_{NL}$, $\phi_{NL}$, $\alpha_{NL}$, $\gamma_{NL}$ and $\delta_{NL}$ can fail assuming that $\delta N$ can be an arbitrary function of $\delta\phi_{i*}$.

The quadra-spectrum and quint-spectrum should obey Maldacena's consistency relations. This consistency relation for general $n$-point correlation has already been discussed in \cite{Huang:2006eha} and \cite{Li:2008gg}.

\section{Multi-brid inflation}

As we have seen in the previous section, with the absence of entropy perturbation, non-Gaussianity is suppressed by slow roll parameters. However, when entropy perturbation is allowed, things get complicated and one might get large quadra-spectrum and quint-spectrum. In this section, we use multi-brid inflation to illustrate the large quadra-spectrum and quint-spectrum in multi-field inflation with entropy perturbation.

We consider a simple model of multi-brid inflation: the two-brid infation with potential
\begin{gather}
  V(\phi_1,\phi_2)=V_0 \exp\left(\alpha_1\phi_1+\alpha_2\phi_2\right)~, \\
  \qquad V_0 \equiv \frac{1}{2}(g_1^2\phi_1^2+g_2^2\phi_2^2)\chi^2+\frac{\lambda}{4}\left( \chi^2-\frac{\sigma^2}{\lambda} \right)^2~.
\end{gather}
The power spectrum and bispectrum of this model is calculated in \cite{Naruko:2008sq}, and the trispectrum is calculated in \cite{Huang:2009xa}. Here we shall generalize the calculation in \cite{Naruko:2008sq, Huang:2009xa} into quadra-spectrum and quint-spectrum.

In this model, inflation ends when $g_1^2\phi_1^2+g_2^2\phi_2^2=\sigma^2$. We parameterize $\phi_1$ and $\phi_2$ at the end of inflation as
\begin{equation}
  \phi_{1f}=\frac{\sigma}{g_1}\cos\gamma~,\quad \phi_{2f}=\frac{\sigma}{g_2}\sin\gamma~.
\end{equation}

The e-folding number before the end of inflation is given as
\begin{equation}\label{efoldinga}
  N(\phi_1,\phi_2)=\frac{1}{2}\ln\left[ \frac{e^{2\phi_1/\alpha_1}+e^{2\phi_2/\alpha_2}}{e^{2\phi_{1f}/\alpha_1} + {e^{2\phi_{2f}/\alpha_2}} } \right] +N_c~.
\end{equation}
where $N_c$ is a correction of e-folding number near the end of inflation, which does not contribute leading order non-Gaussianity when the non-Gaussianity is large. Here $\phi_1 \equiv \phi_{1*}$ and $\phi_2 \equiv \phi_{2*}$ are the value at Hubble crossing. We shall omit the star subscript to avoid clustering of indices, as well as to keep the convention the same as that in the literature.

To take variations with respect to $\phi_1$ and $\phi_2$, one need to note that $\gamma$ should also be written as a function of $\phi_1$ and $\phi_2$. The relation between $\gamma$, $\phi_1$ and $\phi_2$ can be established by the conservation of the angular field component
\begin{equation}\label{cons}
  \frac{\phi_1}{\alpha_1}-\frac{\phi_2}{\alpha_2}=\frac{\phi_{1f}}{\alpha_1}-\frac{\phi_{2f}}{\alpha_2}
  =\frac{\sigma\cos\gamma}{g_1\alpha_1}-\frac{\sigma\sin\gamma}{g_2\alpha_2}~.
\end{equation}
Note that $\phi_1$ and $\phi_2$ have nearly independent and Gaussian perturbations at horizon crossing, which are denoted as $\delta\phi_1$ and $\delta\phi_2$ respectively. Using Eq. \eqref{cons}, the e-folding number \eqref{efoldinga} can be simplified into
\begin{equation}\label{efoldingb}
  N(\phi_1,\phi_2)=(\phi_1-\phi_{1f})/\alpha_1~.
\end{equation}
To expand Eq. \eqref{efoldingb} into fifth order, one have \footnote{Technically, one note that from Eq. \eqref{cons}, the zeroth order quantities satisfy $g\alpha(c_+\cos\gamma+c_-\sin\gamma)=g_1\alpha_1$, which can simplify the calculation.}
\begin{align}
  \delta N_1=\zeta_1 &=\frac{g_1\cos\gamma\delta\phi_1+g_2\sin\gamma\delta\phi_2}{\alpha g c+}~, \nonumber\\
  \delta N_2=\zeta_2 &=\frac{g_1^2g_2^2\left(\alpha_2\delta\phi_1-\alpha_1\delta\phi_2\right)^2}{\sigma g^3\alpha^3 c_+^3}~,\nonumber\\
  \delta N_3=\zeta_3 &=-\frac{3g_1^3g_2^3 c_- \left(\alpha_2\delta\phi_1-\alpha_1\delta\phi_2\right)^3}{\sigma^2 g^4\alpha^4 c_+^5}~,\nonumber\\
  \delta N_4=\zeta_4 &=\frac{3g_1^4 g_2^4 \left( c_+^2+5c_-^2 \right)\left(\alpha_2\delta\phi_1-\alpha_1\delta\phi_2\right)^4}{\sigma^3 g^5 \alpha^5 c_+^7}~,
  \nonumber\\
  \delta N_5=\zeta_5 &=-\frac{15 g_1^5g_2^5 c_-\left(3c_+^2+7c_-^2\right)\left(\alpha_2\delta\phi_1-\alpha_1\delta\phi_2\right)^5}{\sigma^4 g^6\alpha^6 c_+^9}~,
\end{align}
where
\begin{gather}
  \alpha\equiv \sqrt{\alpha_1^2+\alpha_2^2}~,\quad g\equiv \sqrt{g_1^2\cos^2\gamma+g_2^2\sin^2\gamma}~,\nonumber\\
  c_-\equiv \frac{g_1\alpha_1}{g\alpha}\sin\gamma-\frac{g_2\alpha_2}{g\alpha}\cos\gamma~, \quad
  c_+\equiv \frac{g_1\alpha_1}{g\alpha}\cos\gamma+\frac{g_2\alpha_2}{g\alpha}\sin\gamma~,\nonumber\\
  \tilde{c} \equiv \frac{g_2\alpha_1}{g\alpha}\sin\gamma-\frac{g_1\alpha_2}{g\alpha}\cos\gamma~.
\end{gather}
One can either directly calculate the correlation functions or making use of Eq. \eqref{deltaNmpl} to have
\begin{gather}
  f_{NL}=\frac{5\tilde{c}^2g_1^2g_2^2\alpha}{6c_+g^3\sigma}~, \quad
  g_{NL}=\frac{25c_-\tilde{c}^3g_1^3g_2^3\alpha^2}{18c_+^2g^4\sigma^2}~, \quad
  \tau_{NL}=\frac{\tilde{c}^2g_1^4g_2^4\alpha^2}{c_+^2g^6\sigma^2}~, \nonumber\\
  h_{NL}=\frac{125(c_+^2+5c_-^2)\tilde{c}^4g_1^4g_2^4\alpha^3}{216c_+^3g^5\sigma^3}~, \quad
  \theta_{NL}=\frac{3c_-\tilde{c}^3g_1^5g_2^5\alpha^3}{c_+^3g^7\sigma^3}~,
  \quad \phi_{NL}=\frac{\tilde{c}^2g_1^6g_2^6\alpha^3}{c_+^3g^9\sigma^3}~, \nonumber\\
  i_{NL}=\frac{625c_-(3c_+^2+7c_-^2)\tilde{c}^5g_1^5g_2^5\alpha^4}{648c_+^4g^6\sigma^4}~, \quad
  \alpha_{NL}=\frac{3(c_+^2+5c_-^2)\tilde{c}^4g_1^6g_2^6\alpha^4}{c_+^4g^8\sigma^4}~, \quad
  \beta_{NL}=\frac{9c_-^2\tilde{c}^4g_1^6g_2^6\alpha^4}{c_+^4g^8\sigma^4}~, \nonumber\\
  \gamma_{NL}=\frac{3c_-\tilde{c}^3g_1^7g_2^7\alpha^4}{c_+^4g^{10}\sigma^4}~, \quad
  \delta_{NL}=\frac{3c_-\tilde{c}^3g_1^7g_2^7\alpha^4}{c_+^4g^{10}\sigma^4}~, \quad
  \lambda_{NL}=\frac{\tilde{c}^2g_1^8g_2^8\alpha^4}{c_+^4g^{12}\sigma^4}~.
\end{gather}
As shown in \cite{Huang:2009xa}, the parameter $c_+$ can be related with the tensor to scalar ratio $r$ and the spectral index $n_s$ as
\begin{equation}
  c_+^2=\frac{r}{8(1-n_s)}~,
\end{equation}
which should satisfy $c_+^2<0.625$ when $n_s=0.96$ and $r<0.20$. When $r\ll 1$, $c_+ \ll 1$. Thus the quadra-spectrum and quint-spectrum can be large in the multi-brid model. In the $c_+\rightarrow 0$ limit, the order of magnitude of the non-Gaussian estimators are of order
\begin{gather}
  f_{NL}\sim c_+^{-1}~, \quad g_{NL}\sim \tau_{NL}\sim c_+^{-2}~, \quad h_{NL}\sim\theta_{NL}\sim \phi_{NL}\sim c_+^{-3}~,
  \nonumber\\
  i_{NL}\sim\alpha_{NL}\sim\beta_{NL}\sim\gamma_{NL}\sim\delta_{NL}\sim\lambda_{NL}\sim c_+^{-4}~.
\end{gather}
One should also note that in some parameter regions, other parameters could also become large.

\section{The curvaton scenario}

In this section, we calculate the correlation functions up to the sixth order in the curvaton scenario. The calculation is a direct generalization of \cite{Sasaki:2006kq}, in which up to four point correlation functions (corresponding to up to ${\cal O}£¨\zeta_1^3£©$ in the expansion) are calculated. In the calculation, we assume when the curvaton field oscillates, the potential becomes quadratic. We also use sudden decay approximation. One can follow \cite{Enqvist:2005pg} and \cite{Sasaki:2006kq} to generalize the result to go beyond these assumptions.

On super-Hubble scales, the curvature perturbation on uniform density slice can be written as
\begin{equation}\label{deltaN}
  \zeta_i(x)=\delta N(x)+\frac{1}{3}\int_{\bar\rho_i(t)}^{\rho_i(x)}\frac{d\tilde\rho_i}{\tilde\rho_i+\tilde P_i}~,
\end{equation}
where subscript $i$ can denote either inflaton (and the corresponding radiation after inflaton decay), curvaton or a combination of them. As there is no interaction between inflaton and curvaton, the inflaton perturbation $\zeta_r$ and the curvaton perturbation $\zeta_\chi$ are both conserved on super Hubble scales.

Choosing the spatial flat slice, Eq. \eqref{deltaN} for curvaton becomes
\begin{equation}\label{deltaNchi}
  \rho_\chi=\bar\rho_\chi e^{3\zeta_\chi}~.
\end{equation}

As usual, we assume the curvaton could have general slow rolling potential during inflation, while have quadratic potential when the curvaton oscillates. Then the initial amplitude $\chi$ of curvaton oscillation can be written as a function of the curvaton's Hubble exit value $\chi_*$ as $\chi \equiv g(\chi_*)$.

At Hubble exit, the fluctuation of the curvaton field is extremely Gaussian. Thus we can write
\begin{equation}
  \chi_*=\bar\chi_*+\delta_1\chi_*~.
\end{equation}
Correspondingly, the initial amplitude of curvaton oscillation takes the form
\begin{equation}
  g(\chi_*)=g(\bar\chi_*+\delta_1\chi_*)=\bar{g} +\sum_{n=1}^{\infty}\frac{g^{(n)}}{n!}\left(\frac{\delta_1\chi}{g'}\right)^n ~,
\end{equation}
The energy density of curvaton can be written as
\begin{equation}
  \rho_\chi=\frac{1}{2}m^2 g^2 = \frac{1}{2}m^2 \left[ \bar{g} +\sum_{n=1}^{\infty}\frac{g^{(n)}}{n!}\left(\frac{\delta_1\chi}{g'}\right)^n \right]^2~,
\end{equation}
Making use of Eq. \eqref{deltaNchi}, one can relate $\zeta_\chi$ and $\delta\chi$ order by order as
\begin{gather}
  \zeta_{\chi1}=\frac{2}{3}\frac{\delta_1\chi}{\bar\chi}~, \qquad
  \zeta_{\chi2}=\frac{3}{2}\left(-1+\frac{gg''}{g'^{2}}\right)\zeta_{\chi1}^2~, \qquad
  \zeta_{\chi3}=\frac{9}{4}\left(2-3\frac{gg''}{g'^{2}}+\frac{g^2g'''}{g'^{3}}\right)\zeta_{\chi1}^3~, \nonumber\\
  \zeta_{\chi4}=\frac{27}{8}\left(-6+12\frac{gg''}{g'^{2}}-4\frac{g^2g'''}{g'^{3}}-3\frac{g^2g''^2}{g'^{4}}+\frac{g^3 g''''}{g'^{4}}\right)\zeta_{\chi1}^4~,
  \nonumber\\
  \zeta_{\chi5}=\frac{81}{16}\left(24-60\frac{gg''}{g'^{2}}+30\frac{g^2g''^{2}}{g'^{4}}+20\frac{g^2g'''}{g'^{3}}
  -10\frac{g^3g''g'''}{g'^{5}}-5\frac{g^3g''''}{g'^{4}}+\frac{g^4g'''''}{g'^{5}}\right)\zeta_{\chi1}^5~.
\end{gather}
To relate the perturbations $\zeta_\chi$ to $\zeta$, we use the sudden decay approximation. Assuming the curvaton decays on a uniform total density hypersurface $H=\Gamma$, where $\Gamma$ is the curvaton decay rate. Then on this hypersurface we have
\begin{equation}
  \rho_r+\rho_\chi=\bar\rho~.
\end{equation}
Note that on the decay hypersurface, the curvaton is oscillating, thus have equation of state $P_\chi=0$, and the radiation from inflaton has equation of state $P_r=\rho_r/3$. Applying Eq. \eqref{deltaN}, we have
\begin{equation}
  \rho_r=\bar\rho_r e^{4(\zeta_r-\zeta)}~,\qquad \rho_\chi=\bar\rho_\chi e^{3(\zeta_\chi-\zeta)}~.
\end{equation}
Thus on the decay hypersurface we have
\begin{equation}
  (1-\Omega_\chi)e^{4(\zeta_r-\zeta)}+\Omega_\chi e^{3(\zeta_\chi-\zeta)}=1~.
\end{equation}
We consider the standard curvaton scenario, where the primordial inflaton fluctuation $\zeta_r$ can be ignored. In this case, $\zeta$ and $\zeta_\chi$ can be related order by order as
\begin{gather}
  \zeta_1=r\zeta_{\chi1}~,\qquad \frac{\zeta_2}{\zeta_1^2}=\frac{3}{2r}\left(1+\frac{gg''}{g'^2}\right)-r-2~,
  \nonumber\\
  \frac{\zeta_3}{\zeta_1^3}=\frac{9g}{4r^2}\frac{\left(3g'g''+gg'''\right)}{g'^3}-\frac{9}{r}\left(1+\frac{gg''}{g'^2}\right)-\frac{9gg''}{2g'^2}+\frac{1}{2}+10r+3r^2~£¬\label{curvres123}
\end{gather}
\begin{align}
  \frac{\zeta_4}{\zeta_1^4}
  =&\frac{27}{8r^3}\frac{g^2(3g''^2+4g'g'''+gg'''')}{g'^4}-\frac{9}{2r^2}\frac{\left(3g'^4+18gg'^2g''+3g^2g''^2+4g^2g'g'''\right)}{g'^4}
  \nonumber\\
  &+\frac{9}{4r}\frac{\left(17g'^4+2gg'^2g''-3g^2g''^2-4g^2g'g'''\right)}{g'^4}+90+r\left(-50+\frac{27gg''}{g'^2}\right)-70r^2-15r^3~,\label{curvres4}
\end{align}
\begin{align}
  \frac{\zeta_5}{\zeta_1^5}=&
  \frac{81}{16r^4}\frac{g^3\left(10g''g'''+5g'g''''+gg'''''\right)}{g'^5}
  -\frac{135}{4r^3}\frac{g\left( 6g'^3g''+6gg'^2g'''+2g^2g''g'''+9gg'g''^2+9g^2g'g'''' \right)}{g'^5}
  \nonumber\\
  &+\frac{45}{8r^2}\frac{\left(30g'^5+102gg'^3g''+2g^2g'^2g'''-6g^3g''g'''+3g^2g'g''^2-3g^3g'g''''\right)}{g'^5}
  \nonumber\\
  &+\frac{225}{2r}\frac{\left( 3g'^4+12gg'^2g''+3g^2g''^2+2g^2g'g''' \right)}{g'^4}
  +\frac{3}{4}\left( -1713-\frac{1000gg''}{g'^2}+\frac{45g^2\left( 3g''^2+2g'g''' \right)}{g'^4} \right)
  \nonumber\\
  &-350r\left(2+\frac{3gg''}{g'^2}\right)+5r^2(173-\frac{45gg''}{g'^2})+630r^3+105r^4~,\label{curvres5}
\end{align}
where $r\equiv 3\Omega_\chi/(4-\Omega_\chi)$.
Using \eqref{relng}, we have
\begin{gather}
  f_{NL}=\frac{5}{6}\frac{\zeta_2}{\zeta_1^2}~,\qquad g_{NL}=\frac{25}{54}\frac{\zeta_3}{\zeta_1^3}~,\qquad h_{NL}=\frac{125}{648}\frac{\zeta_4}{\zeta_1^4}~,
  \qquad i_{NL}=\frac{125}{1944}\frac{\zeta_5}{\zeta_1^5}~.
\end{gather}
Other non-Gaussian estimators are given in Eq. \eqref{localid}.

In curvaton models, large non-Gaussianity is obtained when $r\rightarrow 0$, which corresponds to the case that curvaton only takes up only a small fraction of total energy density when curvaton decays. In this case non-Gaussianity is large because curvaton has to have larger fluctuations in order to contribute large enough curvature perturbation in uniform total energy density slice. In the $r\rightarrow 0$ limit, the non-Gaussian estimators take the form
\begin{gather}
  f_{NL}\simeq \frac{5}{4r}\left(1+\frac{gg''}{g'^2} \right)~,\quad g_{NL}\simeq \frac{25}{24r^2}\frac{\left(3g'g''+gg'''\right)}{g'^3}~,\nonumber\\
  h_{NL}\simeq \frac{125}{192r^3}\frac{g^2(3g''^2+4g'g'''+gg'''')}{g'^4}~,\quad i_{NL}\simeq \frac{125}{384r^4}\frac{g^3\left(10g''g'''+5g'g''''+gg'''''\right)}{g'^5}~.
\end{gather}

As another special case, when the curvaton potential is always quadratic from Hubble exit to curvaton decay, we have $g''=0,~g'''=0,~g''''=0,~g'''''=0$. In this case, the above result is simplified to be
\begin{gather}
  f_{NL}= \frac{5}{12}\left(\frac{3}{r}-4-2r\right)~,
  \nonumber\\
  g_{NL}= \frac{25}{108}\left( -\frac{18}{r}+1+20r+6r^2 \right)~,
  \nonumber\\
  h_{NL}= -\frac{125}{2592}\left( \frac{54}{r^2}-\frac{153}{r}-360+200r+280r^2+60r^3 \right)~,
  \nonumber\\
  i_{NL}= \frac{125}{7776}\left( \frac{675}{r^2}+\frac{1350}{r}-5139-2800r+3460 r^2+2520 r^3 +420r^4 \right)~.\label{curvresQuad}
\end{gather}
As is well known, in the simplest curvaton model with quadratic potential (corresponding to Eq. \eqref{curvresQuad}), in the $r\ll 1$ limit, $g_{NL}$ is of order $1/r$, the same scaling as $f_{NL}$. Such a $g_{NL}$ as small as $g_{NL}\ll f_{NL}^2$ is much more difficult to be measured than $f_{NL}$. However, $h_{NL}$ scales as $1/r^2$, we have $g_{NL}^{3/2}\ll h_{NL} \ll f_{NL}^3$. Thus the situation for measuring $h_{NL}$ should be better than $g_{NL}$, although more difficult than $f_{NL}$. For similar reason, $i_{NL}$ is much more difficult to be measured than $h_{NL}$. However, in the more general case , where the curvaton potential is not quadratic during inflation (corresponding to Eqs. \eqref{curvres123}, \eqref{curvres4},\eqref{curvres5}), there is no such hierarchy, and all terms are equally likely to be measured. The counter part of this effect for trispectrum is investigated in \cite{Enqvist:2005pg}.

\section{Non-local non-Gaussianities}
In this section, we would like to go beyond the local ansatz, which leads to the shape function other than those defined in Eqs. \eqref{shapeQ} and \eqref{shapeU}. We illustrate the non-local shape using DBI inflaton. The complete calculation is very lengthy, with a great number of terms, which is beyond the scope of the current paper. However, from the full calculation of bispectrum and trispectrum, we get the experience that all equilateral shape functions are similar. More technically speaking, the shape functions of the bispectrum generally have a large component of the standard equilateral shape, and have a very small mixing of orthotropic shape. Similar phenomenon is also observed in trispectrum. Thus we could expect that we are able to extract some shape information of the quadra-spectrum and the quint-spectrum, even when the calculation is not yet complete.

Before a calculation, we want to estimate the order of magnitude for the result. When the kinetic term of inflaton is generalized, the sound speed for perturbation is generally different from unity. By power counting \cite{Huang:2006eha}, one expects $n$-point correlation function ($n\geq 3$) scales as
\begin{equation}
  \langle \zeta^n \rangle \propto P_\zeta^{n-1}/ c_s^{2n-4}~,
\end{equation}
where $P_\zeta$ is the dimensionless power spectrum, and $c_s$ is the sound speed for perturbation. Thus one have
\begin{equation}
  h_{NL}\propto c_s^{-6} \sim f_{NL}^3 ~,\qquad i_{NL}\propto c_s^{-8} \sim f_{NL}^4~.
\end{equation}
This behavior is similar with the curvaton model with non-quadratic term.

We perform the calculation in spatial flat gauge. In the $c_s \ll 1$ limit, the 5th and 6th order action of DBI inflation takes the form
\begin{gather}
  S_5=\int d^4 x \left\{ \frac{7a^3\dot{\delta\phi}^5}{8c_s^9\dot\phi^3}
  -\frac{5a(\partial_i\delta\phi\partial_i\delta\phi)\dot{\delta\phi}^3}{4c_s^7\dot\phi^3}
  +\frac{3(\partial_i\delta\phi\partial_i\delta\phi)^2\dot{\delta\phi}}{8ac_s^5\dot\phi^3}
   \right\}~. \label{s5dbi}\\
  S_6=\int d^4 x \left\{ \frac{21a^3\dot{\delta\phi}^6}{16c_s^{11}\dot\phi^4}-
  \frac{35a(\partial_i\delta\phi\partial_i\delta\phi) \dot{\delta\phi}^4}{16c_s^9\dot\phi^4}+\frac{15(\partial_i\delta\phi\partial_i\delta\phi)^2\dot{\delta\phi}^2}{16ac_s^7\dot\phi^4}
  -\frac{(\partial_i\delta\phi\partial_i\delta\phi)^3}{16a^3c_s^5\dot\phi^4} \right\}~. \label{s6dbi}
\end{gather}

In the interaction picture, the field $\delta\phi$ can be expanded in terms of creation and annihilation operators as
\begin{equation}
  \delta\phi^I(\tau,\bk)=u(\tau,\bk)a(\bk)+u^*(\tau,-\bk)a^\dagger(-\bk)~,
\end{equation}
The mode function $u(\tau,\bk)$ satisfies the classical equation of motion
\begin{equation}
  u_k=\frac{H}{\sqrt{2k^3}}\left(1+ikc_s\tau\right)e^{-ikc_s\tau}~,\quad u'_k=\frac{H\sqrt{k}}{\sqrt{2}}c_s^2\tau e^{-ikc_s\tau}~.
\end{equation}

We are fully aware that there are non-trivial transformation from Lagrangian to Hamiltonian, and one also need to calculate diagrams with scalar propagation. However, to have some feelings of the amplitude and the shape function, we only calculate the first term each for \eqref{s5dbi} and \eqref{s6dbi}, disregarding all the complexities.
\footnote{From the experience of bispectrum, the first term might be canceled by other terms in DBI inflation. However, the first term is the simplest, while hopefully carrying the general information of the shape functions. In the case of trispectrum, the corresponding term is chosen as the representive shape. Thus we still choose to calculate the first term. We might, with a little more effort, calculate all the terms listed in the action. However, an incomplete result makes no much improvement compared with a more incomplete result for illustration purpose.}

It is convent to use the commutator form of the in-in formalism
\begin{equation}
  \langle \zeta(t)^n \rangle = -i \int_{t_0}^{t}dt' \langle  [\zeta_I(t)^n, H_{\rm int}(t')] \rangle~.
\end{equation}
For the quadra-spectrum, we have
\begin{align}
  \langle \zeta(\tau,\bk_1)\zeta(\tau,\bk_2)\zeta(\tau,\bk_3)\zeta(\tau,\bk_4)\zeta(\tau,\bk_5) \rangle \supset
  -\frac{4725P_\zeta^4 (2\pi)^{11}\delta^3(\sum_i\bk_i)}{c_s^6 k_1k_2k_3k_4k_5K_5^7}~,
\end{align}
where $K_5\equiv k_1+k_2+k_3+k_4+k_5$. For the quint-spectrum, we have
\begin{align}
  \langle \zeta(\tau,\bk_1)\zeta(\tau,\bk_2)\zeta(\tau,\bk_3)\zeta(\tau,\bk_4)\zeta(\tau,\bk_5)\zeta(\tau,\bk_6) \rangle \supset
  -\frac{1190770P_\zeta^5(2\pi)^{13}\delta^3(\sum_i\bk_i)}{c_s^8 k_1k_2k_3k_4k_5k_6K_6^9}~,
\end{align}
where $K_6\equiv k_1+k_2+k_3+k_4+k_5+k_6$. Note that the quarda-spectrum and the quint-spectrum are proportional to $c_s^{-6}$ and $c_s^{-8}$ respectively, as expected. The numerical coefficients are also large because a large number of permutations.

\section{Conclusion and discussion}

To conclude, we discussed quadra-spectrum and quint-spectrum in inflation and curvaton models. In inflation models, these correlations are in general suppressed by slow roll parameters, with the exception of oscillation, nontrivial entropy perturbation or generalized kinetic terms. In curvaton models, especially in curvaton models with potential other than quadratic terms during inflation, the quadra-spectrum and quint-spectrum can get large.

The importance of quadra-spectrum and quint-spectrum depends on the experimental ability to probe them. In this paper, we simply assume the quadra-spectrum and the quint-spectrum are ``large'' when $h_{NL}\geq f_{NL}^3$, and $i_{NL}\geq f_{NL}^4$, respectively. However, this estimation need to be verified or improved by data analysis. For CMB data, we might naively expect that the signal to noise ratio of quadra-spectrum and quint-spectrum are direct generalizations of Eq. \eqref{sn} like
\begin{equation}
  (S/N)_T\sim P_\zeta^{3/2}f_{NL}^3 l^3~, \quad (S/N)_U\sim P_\zeta^{2}f_{NL}^4 l^4~,
\end{equation}
which supports our estimation when one can reach $l\sim$thousands in Planck, with the assumption that $f_{NL}\geq 10$. Moreover, one can hope that with the help of the future LSS experiments, the data points are available in three dimensions instead of two. In this case, the number of available data points increases, which favors higher order correlation functions. Another possibility which favors higher order correlations is that in some cases, the quadra-spectrum and quint-spectrum are indeed much larger than $f_{NL}^3$ and $f_{NL}^4$ respectively. This possibility can be achieved by tuning the parameters in the models we have discussed, or more naturally achieved by a possibility that we shall discuss as follows.

In quasi-single field inflation, which is an intermediate case between single field and multi-field inflation, the quadra-spectrum and quint-spectrum are large, and potentially be promising in probing non-Gaussianities. This is because in quasi-single field inflation, the non-Gaussianity in the isocurvature direction need to be projected onto inflation direction using transfer vertices. Using the result of \cite{Chen:2009we}, the non-Gaussian estimator $f_{NL}$ in a turning trajectory model of quasi-single field inflation is of order
\begin{equation}
  f_{NL}\sim P_\zeta^{-1/2}\left(\dot\theta/H\right)^3 \left(V'''/H\right)~,
\end{equation}
where $\dot\theta/H$ denotes the turning angle of inflation trajectory during one Hubble time, and $V'''$ denote the interaction in the isocurvature direction. To use perturbation theory in calculation, we need $\left(\dot\theta/H\right)^2\ll 1$. As in \cite{Chen:2009we}, one can estimate the order of magnitude of quadra-spectrum and quint-spectrum as
\begin{gather}
  h_{NL}\sim P_\zeta^{-3/2}\left(\dot\theta/H\right)^5 \left(V'''/H\right)^3 \sim (\dot\theta/H)^{-4} f_{NL}^3~,\\
  i_{NL}\sim P_\zeta^{-2}\left(\dot\theta/H\right)^6 \left(V'''/H\right)^4 \sim (\dot\theta/H)^{-6} f_{NL}^4~.
\end{gather}
If the transfer vertex can be calculated perturbatively, then the factors $(\dot\theta/H)^{-4}$ and $(\dot\theta/H)^{-6}$ are much greater than 1. Thus quadra-spectrum and quint-spectrum may play interesting roles in quasi-single field inflation. However, as an explicit calculation of bispectrum is already messy in quasi-single field inflation, one had better improve the calculation technique in calculating quadra-spectrum and quint-spectrum for quasi-single field inflation.

As an extremal example, when $V'''\sim H$, the isocurvature direction is completely non-Gaussian. In this case, if $f_{NL}$ is intermediately large, say $f_{NL}=10$, we will have $h_{NL}\sim f_{NL}^7$, instead of the cubic power.

The discussion on quasi-single field inflation can be made more general. In multi-field inflation as well, non-Gaussianity might be created in isocurvature directions and be converted to inflation direction. Sometimes one ignores this effect not only because of simplicity, but also because transfer vertices are involved, which might suppress the correlation function. However, quadra-spectrum and quint-spectrum need less transfer vertices compared with $f_{NL}^3$ and $f_{NL}^4$ respectively. Thus for quadra-spectrum and quint-spectrum for multi-field inflation, isocurvature perturbations also become more important.

\medskip
\section*{Acknowledgments}
We thank Xingang Chen, Miao Li and Eiichiro Komatsu for discussions. YW was supported by NSERC and an IPP postdoctoral fellowship. CL was supported by a NSFC grant No.10535060/A050207, a NSFC grant No.10975172, a NSFC group grant No.10821504 and Ministry of
Science and Technology 973 program under grant No.2007CB815401.

\end{document}